\documentstyle[prl,aps]{revtex}
\input{psfig}
\draft
\begin{document}
\twocolumn[\hsize\textwidth\columnwidth\hsize\csname@twocolumnfalse\endcsname
\title{
Driven Vortex States and Relaxation in Single Crystal $YBa_2Cu_4O_8$} 
\author{
V. Metlushko, U. Welp, I. Aranson, S. Scheidl\cite{sh}, V. M. Vinokur, G. W. Crabtree} 
\address{
Materials Science Division, Argonne National Laboratory, Argonne, IL 60439} 
\author{K. Rogacki\cite{rg}, B. Dabrowski} 
\address{Department of Physics, Northern Illinois University, DeKalb, IL 60115} 

\date{\today}
\maketitle
\begin{abstract}
The vortex response to various ac-drives has been studied in high quality $YBa_2Cu_4O_8$ single crystals.  Well within the vortex solid phase a re-entrant resistive state has been observed that is characterized by long relaxation times.  Pulse current response and IV characteristics reveal that in this state the driven vortex system relaxes in a way to increase t he effective pinning force.  A  phenomenological model accounts for the observed features, in particular the clock-wise hysteresis of the IV-curves and the increase of the apparent critical current with applied current amplitude.
\end{abstract} 
\narrowtext
\vskip1pc]

The behavior of the vortex system in type-II superconductors in the presence of an applied drive current has been the subject of intense recent research.  Of particular importance for technical applications of superconductors as well as for their fundamental understanding is the mechanism of depinning at a critical drive current which is believed to involve tearing of the vortex lattice and plastic flow \cite{1,2}. Strong history effects and current induced transitions between high and weak pinning metastable states \cite{3} have been reported.  New relaxation phenomena have recently been observed in $NbSe_2$ at sub-critical alternating pulse drives \cite{4} a full theoretical understanding of which has not yet emerged.            

Here we present magneto-transport measurements on single crystal $YBa_2Cu_4O_8$ using various pulse and ac-drives.  In the vortex solid state we observe a novel re-entrant, strongly non-linear response that is characterized by long relaxation times.  It is induced by ac-drives with amplitudes well below the dc-critical current.  In contrast to the usually observed current-induced decrease in the pinning force, in this region of the phase diagram the driven vortex system relaxes in a way to increase the effective pinning force.  A simple phenomenological model accounts for the principal observed features, in particular the clock-wise hysteresis of the IV-curves and the increase of the apparent critical current with applied current amplitude.  We suggest that current-induced vortex entanglement is a possible mechanism underlying this behavior.  

The $YBa_2Cu_4O_8$ crystal used in this study has the dimensions $a \times b \times c  = 0.26 \times 0.85 \times 0.038$  mm$^3$ and was grown in a self-flux method under high oxygen pressure, $T_c=80$ K  \cite{5}.  The occurrence of the various liquid and solid vortex phases was inferred from the temperature dependence of the resistivity for fields applied along the c-axis as shown in Fig. 1. Similar results have recently been reported by Qiu et al. \cite{19}.   At a resistivity level of roughly 15\% of the normal state resistivity at $T_c$ a sudden drop is observed which smears out in high fields and that coincides with the onset of non-Ohmic transport as shown in the bottom inset of Fig. 1.  This behavior is reminiscent to that found for un-twinned $YBa_2Cu_3O_{7-d}$ \cite{6}, and in analogy we identify the onset of non-Ohmic behavior with the freezing/melting transition.  The melting line deduced from these data is well described by $H_m(T) = 32 (1 - T/T_c)^{1.4}$  and shown in comparison with the melting line of $YBa_2Cu_3O_{7-d}$ \cite{8} in the top inset of Fig. 1.  Also shown is the scaling $H_m(T)/H_{cr}$  with $H_{cr} = \Phi_0/(\gamma s)^2$ and $\gamma$ the anisotropy parameter and $s$ the $CuO_2$-layer spacing, which is expected for Josephson-coupled vortices \cite{7}.  The melting lines of both materials can be approximately collapsed onto one curve.  The values of $\gamma$ = 7.5 and 12.5 have been used for $YBa_2Cu_3O_{7-d}$ \cite{8} and  $YBa_2Cu_4O_8$ \cite{9}.

Fig. 2 shows the field dependence of the resistivity at 70 K near the tail of the melting transition measured with a sinusoidal current with amplitude of 20 mA at various frequencies. The data are taken for increasing field after the sample has been zero-field cooled.  Pronounced hysteretic effects between field cooled and zero-field cooled measurements will be described elsewhere \cite{10}.  Well within the vortex solid state we observe a strongly frequency dependent re-entrant resistive state.  The re-entrant behavior is reminiscent of the peak-effect phenomenon \cite{11}, and previous magnetic measurements \cite{9,12,13} at temperatures below 60 K have shown pronounced peak-effects superimposed on weak bulk pinning.  However, the state observed here appears only for ac-drives: with dc-currents up to 60 mA (the highest current that can be passed without excessive contact heating) no voltage response was detected.  The frequency dependence of the resistivity at fixed fields is shown in the inset of Fig. 2.  For fields in the re-entrant region a well defined cross-over from a strong increase roughly proportional to $\omega^{3/4}$ at frequencies below 500 Hz to a weaker $\omega^{1/4}$ or possibly logarithmic dependence at frequencies above 500 Hz is observed.  This defines a characteristic time scale of about 2 $msec$.  In contrast, the melting transition itself is essentially frequency independent in the frequency range studied here in agreement with results on $NbSe_2$ \cite{14} and $YBa_2Cu_3O_{7-d}$ \cite{15}.  The long time constants in the re-entrant region indicate complex dynamics possibly involving glassy relaxation and/or plastic deformations of the vortex system.

The response to current pulses in the form of a symmetric square wave is shown in Fig. 3.  These data as well as those in Fig. 4 are obtained from an average over a large number of cycles using a digital storage oscilloscope with 7 kHz bandwidth.  They therefore represent the periodic response, the initial response on the first few cycles cannot be resolved.  In the tail of the melting transition at 1.2 T the expected instantaneous voltage response in the form of a symmetric square wave is observed whereas in the re-entrant resistive region at 0.4 T the voltage, after an instantaneous response on each current reversal, decays rapidly to essentially zero.  The recurrence of the voltage response on every cycle shows that these data do not represent a current induced transition from a weak pinning metastable state into a high pinning state as discussed in Ref. \cite{3}.  The data at 92 Hz and 230 Hz indicate that the longer the time interval of constant current the further the voltage decays.  The characteristic decay time of about 2 msec is consistent with the data in Fig. 2, and  is independent of the applied current amplitude ranging from 15 to 45 mA.  This time dependence accounts for the strong frequency dependence shown in Fig. 2 and the absence of a response under dc-drives.  Assuming that the 
data  shown in Fig. 3 represents the motion of the entire vortex system (this excludes channel flow) then an average travel distance of 0.7 $\mu$m can be estimated which corresponds to about 10 vortex lattice constants.  Very similar symmetric pulse responses have recently been reported for $NbSe_2$ \cite{4}.  An oscillatory vortex response with periods much larger than those of the drive current has recently been observed in $YBa_2Cu_3O_{7-d}$ below the melting line \cite{21}.  In contrast, the vortex response shown in Figs. 3 and 4 occurs on the time scale of the drive and represents therefore a different dynamic state.  

The IV-characteristics of $YBa_2Cu_4O_8$ taken near the maximum of the re-entrant region with sinusoidal currents are shown in the main panels of Figs 4a and 4b for 70 K and 75 K.  The highly non-linear voltage response (see right half of Fig. 4a) is characterized by a sizable hysteretic (that is, out-of-phase) component which has unusual clock-wise rotation.  On the left half of Fig. 4a the in-phase component, $V_1$, and the out-of-phase component, $V_2$, are shown separately.  At current amplitudes around 15 mA the voltage response rises quickly.  However, with further increasing current amplitude the voltage response does not follow the initial IV-characteristics but a characteristic that is progressively parallel shifted to higher current values: the apparent critical current increases with increasing amplitude of the applied current.  At 70 K this unusual behavior persists to the highest attainable amplitudes.  The progression of the in-phase component, $V_1$, of the IV-curves taken at 75 K is shown in Fig. 4b.  The out-of-phase component at this temperature is strongly reduced as compared to the data in Fig. 4a.  At 75 K the usual IV-behavior is recovered for current amplitudes exceeding 40 mA.  The apparent critical current as determined from the evolution of $V_1$ at 70 K and 75 K and using a voltage criterion of 50 nV is shown in the lower inset of Fig. 4b.  Except for current amplitudes close to the threshold value the apparent critical currents for increasing and decreasing amplitude coincide and vary almost linearly with applied current amplitude.

The response of the vortex system to an external electromagnetic perturbation can be described on a macroscopic level by the non-linear diffusion equation for the local current density \cite{16}:
$\mu_0 d{\bf j}({\bf r},t)/dt = {\bf \nabla}^2(\rho (j) {\bf j}({\bf r},t))$.  Numerical integration \cite{20} of this equation for current flow in a slap with a thickness much less than its width and non-linear resistivities of the form $\rho (j) = \rho _0 (j/j_c)^n$ as well as 
$\rho (j) = \rho _0 exp(-j_c/j)$ indicates that the results presented here can not be accounted for through current redistributions due to skin effect or inhomogenous critical states. 
The data shown in Figs. 3 and 4 represent a novel dynamic state in which the vortex system adjusts itself under the influence of a driving force in a way to enhance the effective pinning force.  We  model this relaxation  behavior through a phenomenological extension of the  hysteretic elastic creep theory \cite{17}.  Within the elastic creep region the dominant current dependence in the IV characteristic is given by 
$E = \rho  j \exp(-j_{th}/|j|)$  where  $j_{th}  = U_c j_c /k_B T$.  Here $j_c$ is the critical current density, 
$U_c$ is the typical pinning energy, and $\rho$ is the resistivity scale.  We adopt a value one for the glassy exponent and model the critical current relaxation by the simplest "equation of motion" for $j_{th}$:  $d j_{th}/dt = -\eta  j_{th} + g (|j|)$, where the constant $\eta$  characterizes the rate of relaxation of $j_{th}$ to its equilibrium value and the function $g$ describes the effect of the applied current.  In leading order we can expand $g = g_0 + \delta  |j| + $... For a sinusoidal variation $j = j_0 \sin(\omega t)$,  $j_{th}(t)$  contains a periodic contribution and, in addition to $g_0/\eta$  a constant off-set that is proportional to $\delta j_0$.  The parameters $\rho , \eta, g_0$  and $\delta $ were determined from fitting the data in Figs. 3 and 4 resulting in $\eta$ = 400 Hz, $\delta$  = 8000 Hz, $g_0$  = 0 and $\rho$  = 4.7 m$\Omega$ cm.  The fits are shown in the figures.  The clock-wise hysteresis, the parallel shift of the IV-curves with current amplitude and the voltage decay for square-pulse currents, can be reproduced.  In this model the clock-wise hysteresis arises because the periodic part of $j_{th}(t)$ reaches its maximum value with a delay with respect to the extrema of $j(t)$.  This delay is determined by the relaxation parameter $\eta$ and causes a smaller voltage on the decreasing branch of $|j(t)|$ as compared to the increasing branch.  Similarly, the shift of the IV-curves is caused by the constant off-set in $j_{th}(t)$ that increases linearly with applied current amplitude.  This linear dependence is in fact observed in the apparent critical current (lower inset, Fig. 4b).  We note that the apparent critical current extrapolates essentially to zero, consistent with $g_0=0$.  The value of $\eta$ = 400 Hz sets the time scale of the vortex response and is in good agreement with the frequency dependence shown in Fig. 2 and the pulse-current response (see Fig. 3).  However, the rapid voltage decay at short times (that is, high frequencies) in the pulse response as well as the threshold behaviors at the onset of the unusual IV-behavior and at the recovery of the usual behavior are not captured.  This discrepancy may arise from the assumptions underlying this model, in particular, the linear expansion of $g(|j|)$, and a more complicated form of this function is required.

We suggest  that vortex entanglement \cite{18} is a possible microscopic mechanism underlying the observed behavior.  An applied drive current can cause the forced entanglement of vortex lines which will lead to a substantial suppression of vortex creep in much the same way as during work hardening the increasing number of dislocations effectively leads to a blockage of their motion.  This argument is in qualitative agreement with the simulations of vortex motion in three-dimensional layered superconductors modeled by the time-dependent Ginzburg-Landau equation.  
As a measure of entanglement we calculated the diffusion of vortex lines under the influence of a drive current, $\langle | r(z) - r(z^\prime )|^2 \rangle = D |z - z^\prime |$.  Here, $r(z)$ is the position of the vortex line,$ z$ the coordinate along the field direction and $D$ the "diffusion" coefficient.  Our simulations indicate a sharp increase of the entanglement near the onset of vortex motion, see top inset of Fig. 4b.

In conclusion, the states of vortex dynamics in $YBa_2Cu_4O_8$ have been investigated using magneto-transport measurements.  The melting line $H_m(T)$  separating the vortex liquid and vortex solid states has been determined from the onset of 
non-Ohmic behavior.  $H_m$  shows good scaling according to $H_m  \sim \gamma^{-2 } $ with the results of $YBa_2Cu_3O_{7-d}$.  Well within the solid state a re-entrant strongly current and frequency dependent resistive state is observed.  Pulse-current response and IV-characteristics reveal that in this region of the phase diagram the driven vortex system relaxes in a way to increase the effective pinning force.  A simple phenomenological model accounts for the observed clock-wise hysteresis of the IV-curves and the increase of the apparent critical current with applied current amplitude.  We argue that forced vortex entanglement due to the drive current is a possible microscopic mechanism underlying this dynamic vortex state.

This work was supported by the US DOE, BES - Materials Sciences under contract \#W-31-109-ENG-38 (V.M., U.W., I.A., V.M.V., G.W.C.),  by Deutsche Forschungsgemeinschaft Projekt SFB 341 and grant SCHE/513/2-1 (S.S.) and by the NSF STCS  under contract \#DMR 91-20000 (I.A., K.R., B.D.).  We acknowledge helpful discussions with A. Koshelev and V. Vlasko-Vlasov.

\leftline{\psfig{figure=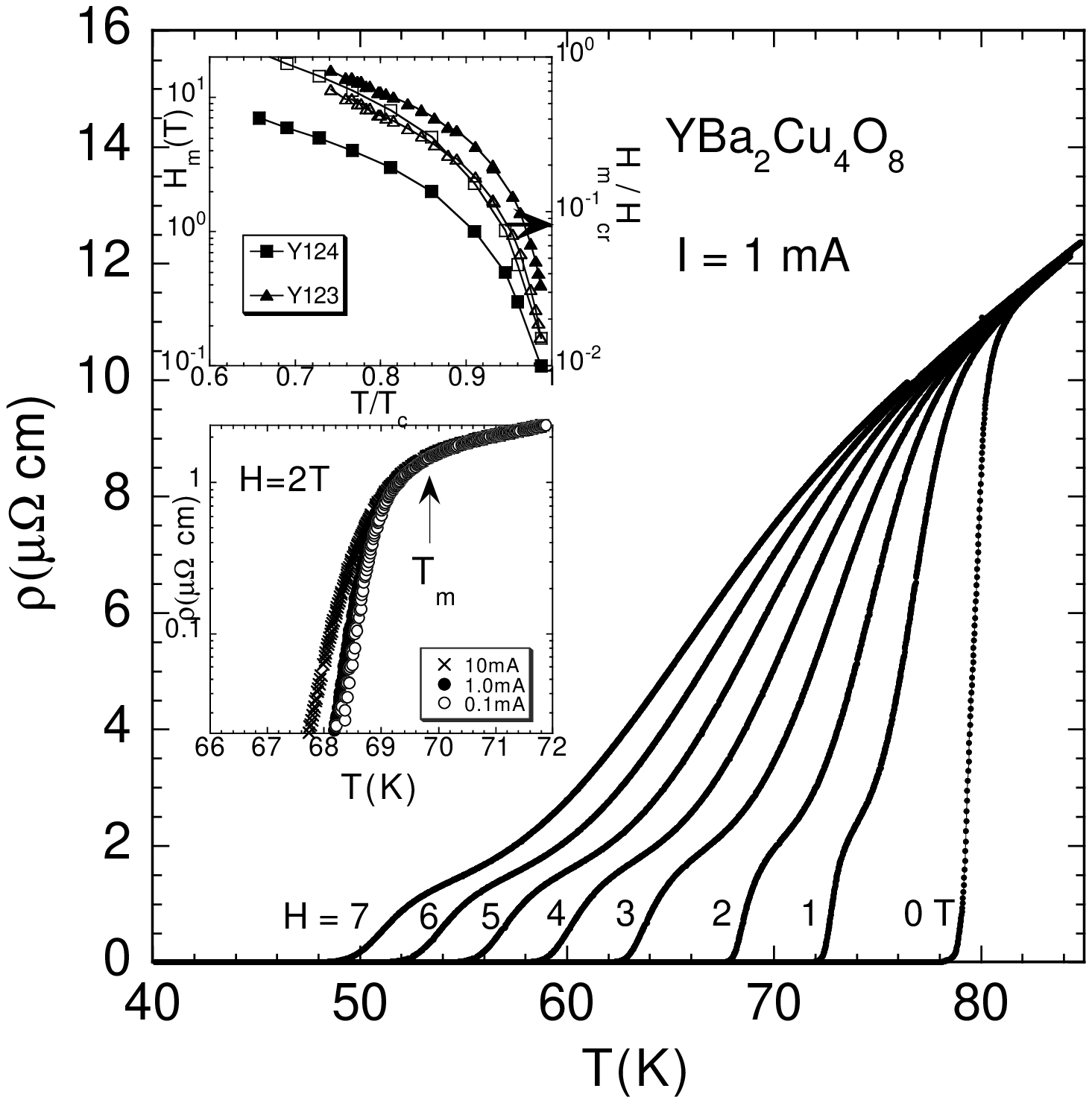,height=3.in}}
\begin{figure}
\caption{
Temperature dependence 
of the resistivity of $YBa_2Cu_4O_8$ in various fields 
applied parallel to the c-axis. 
Bottom inset: Onset of non-Ohmic behavior at 
the melting transition.  Top inset: Comparison of the melting lines
$ H_m(T)$ (solid symbols) and scaled melting lines $H_m/H_{cr}$
(open symbols) of $YBa_2Cu_3O_{7-d}$ and  $YBa_2Cu_4O_8$.
}
\label{fig1}
\end{figure}
\leftline{    \psfig{figure=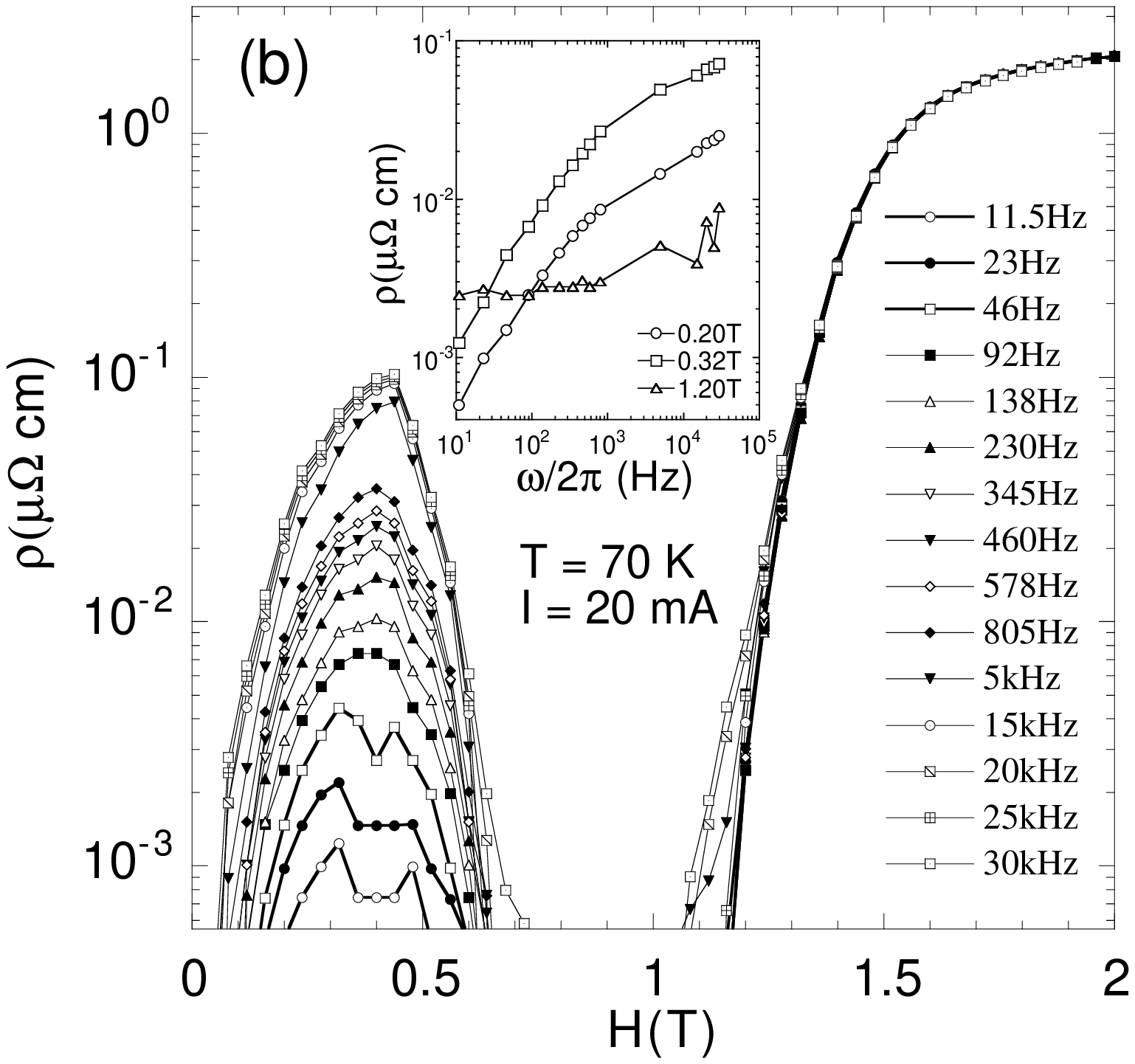,height=3in} }
\begin{figure}
\caption{ 
Field dependence of the in-phase resistivity at 70 K measured with ac-currents of various frequencies.  
Inset: Frequency dependence of the resistivity at two fields in the re-entrant region and in 
the tail of the melting transition.
}
\label{fig2}
\end{figure} 
\leftline{\psfig{figure=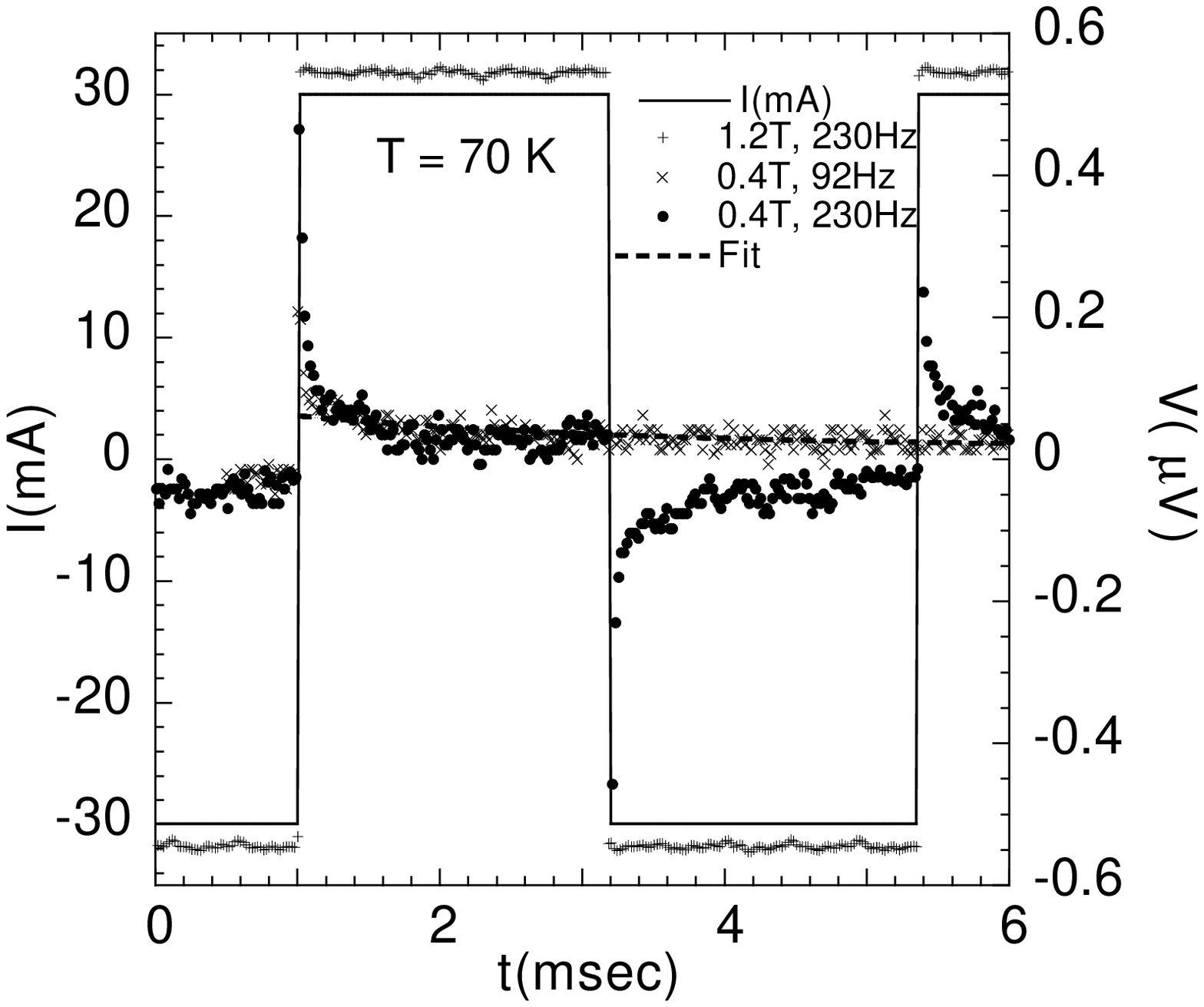,height=3.in}    }

\begin{figure}
\caption{
Voltage response to currents in the form of symmetric square-pulses at 70 K for magnetic fields in the re-entrant region and in the tail of the melting transition.  The heavy broken line is a fit according to the phenomenological model.
}
\label{fig3} 
\end{figure}

\leftline{ \psfig{figure=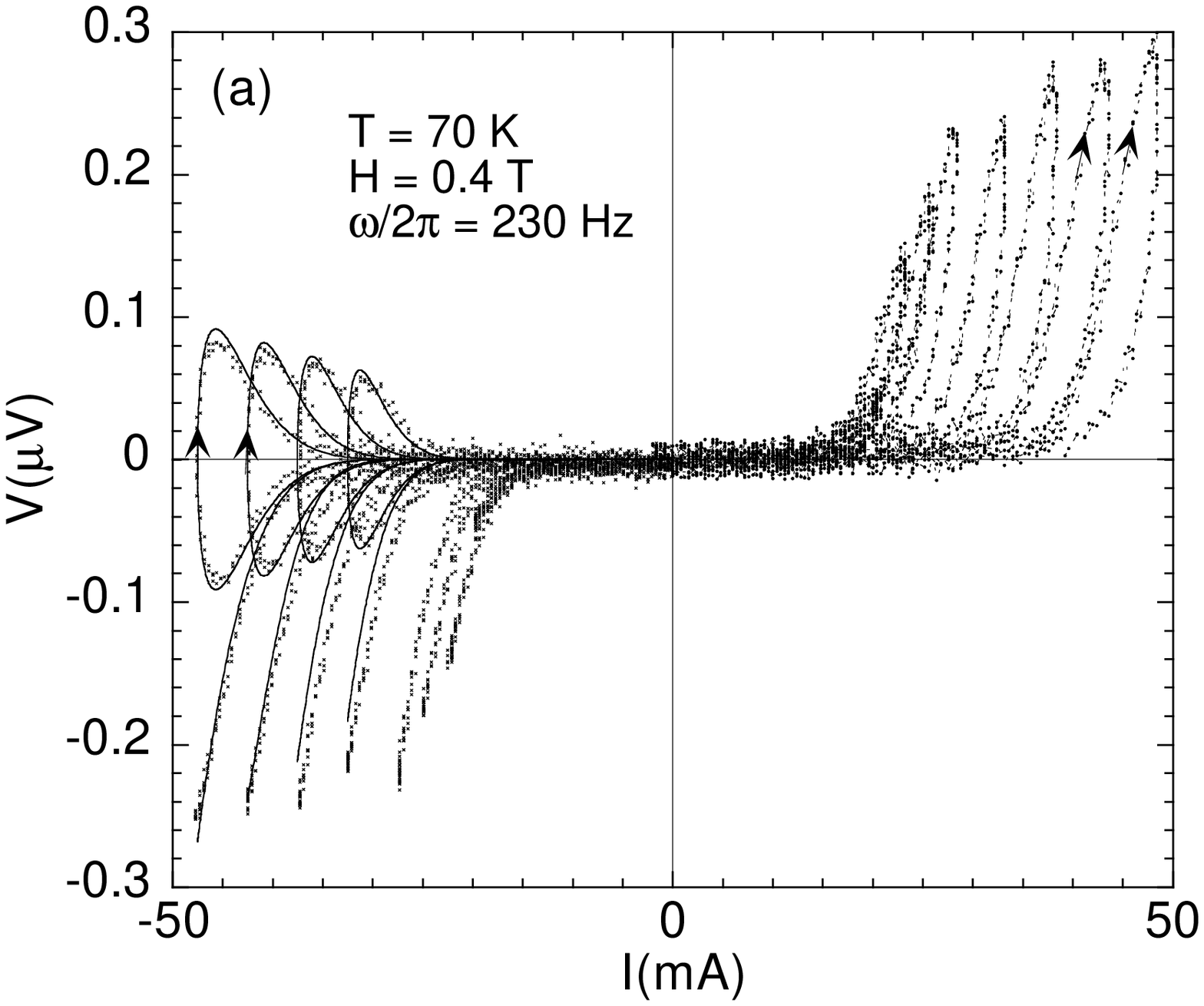,height=2.7225in} } 
\leftline{    \psfig{figure=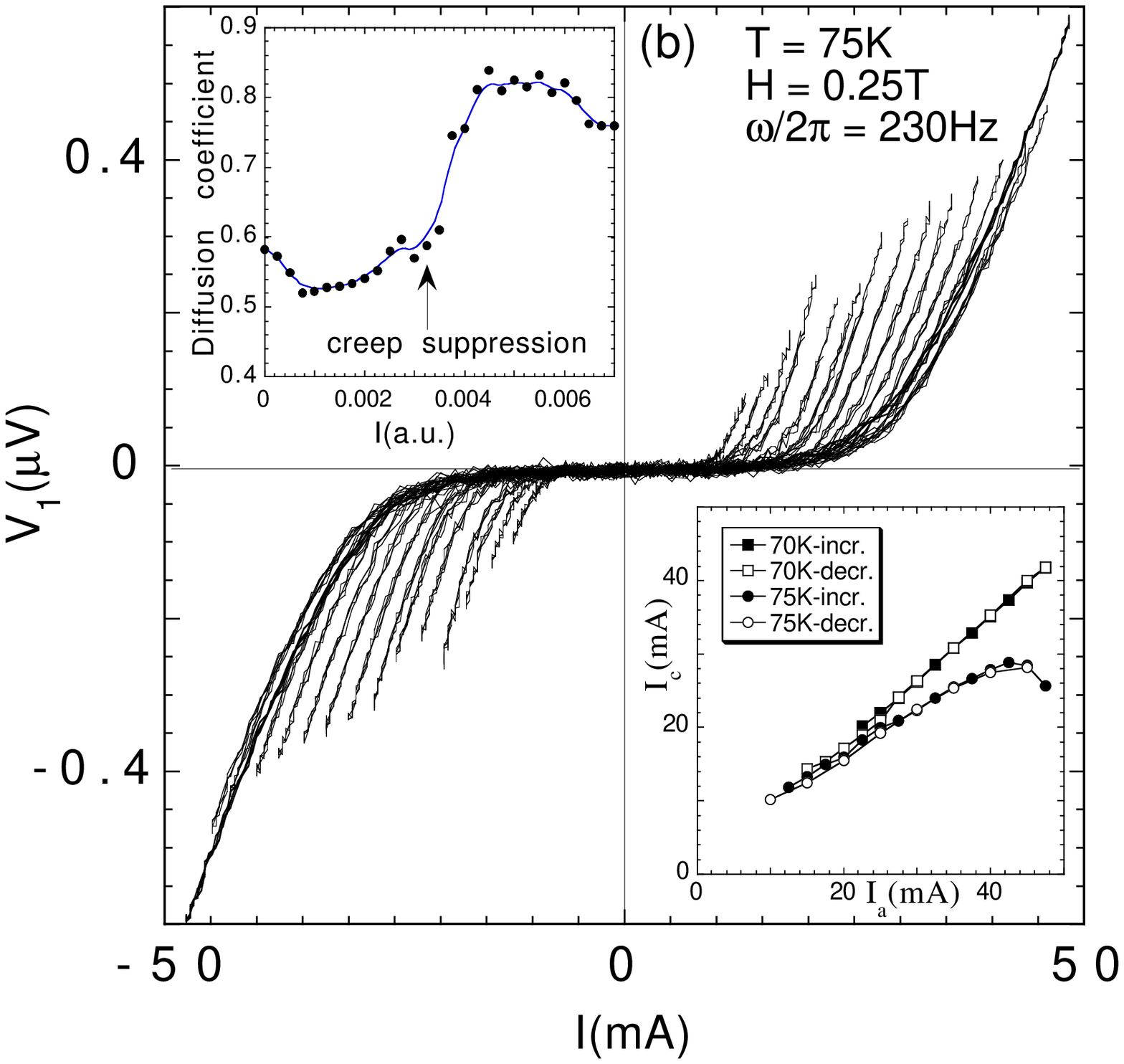,height=3.01in} }
\begin{figure}
\caption{
(a) Current-voltage characteristics at 70 K and 0.4 T for various 
amplitudes of the applied sinusoidal current.  The right half of the figure shows  the total voltage response.  The clock-wise sense of rotation is indicated by the arrows.  
In the left of the figure the in-phase and out-of-phase component are shown separately.  The lines are fits according to the phenomenological model; 
(b) Current-voltage characteristics for the in-phase component $V_1$
 at 75 K and 0.25 T.    Bottom inset: Amplitude dependence of the apparent critical current at 70 K and 75 K as 
determined from a 50 nV criterion.  Top inset: Current dependence of the vortex diffusion coefficient.
}
\label{fig4} 
\end{figure} 
\end{document}